\documentclass{emulateapj}

\newcommand{\MyPulsar}{PSR~J1744$-$3922}
\newcommand{\mypulsar}{PSR~J1744$-$3922}
\newcommand{\Msun}{\ensuremath{\textrm{M}_{\sun}}}

\slugcomment{Accepted for publication in ApJ, 2007 February 13}

\shorttitle{The Unusual Binary Pulsar PSR~J1744$-$3922}
\shortauthors{Breton et al.}

\begin{document}

\title{The Unusual Binary Pulsar PSR~J1744$-$3922: Radio Flux Variability,
Near-infrared Observation and Evolution}

\author{R. P. Breton\altaffilmark{1}, M. S. E. Roberts\altaffilmark{2}, S. M.
Ransom\altaffilmark{3}, V. M. Kaspi\altaffilmark{1}, M.
Durant\altaffilmark{4}, P. Bergeron\altaffilmark{5} and A. J.
Faulkner\altaffilmark{6}}

\altaffiltext{1}{Department of Physics, Rutherford Physics Building, McGill
University, Montreal, QC H3A 2T8, Canada; bretonr@physics.mcgill.ca}
\altaffiltext{2}{Eureka Scientific, Inc., 2452 Delmer Street Suite 100
Oakland, CA 94602-3017}
\altaffiltext{3}{National Radio Astronomy Observatory, 520 Edgemont Road,
Charlottesville, VA 22903}
\altaffiltext{4}{Department of Astronomy and Astrophysics, University of
Toronto, 60 St. George Street, Toronto, ON M5S 3H8, Canada}
\altaffiltext{5}{D\'epartement de Physique, Universit\'e de Montr\'eal, C.P.
6128, Succ. Centre-Ville, Montr\'eal, QC H3C 3J7, Canada}
\altaffiltext{6}{University of Manchester, Jodrell Bank Observatory,
Macclesfield, Cheshire, UK SK11 9DL}

\begin{abstract}
\MyPulsar\ is a binary pulsar exhibiting highly variable pulsed radio
emission. We report on a statistical multi-frequency study of the pulsed
radio flux variability which suggests that this phenomenon is extrinsic to
the pulsar and possibly tied to the companion, although not strongly
correlated with orbital phase. The pulsar has an unusual combination of
characteristics compared to typical recycled pulsars: a long spin period
(172\,ms); a relatively high magnetic field strength ($1.7 \times
10^{10}$\,G); a very circular, compact orbit of 4.6 hours; and a low-mass
companion (0.08\,\Msun). These spin and orbital properties are likely
inconsistent with standard evolutionary models. We find similarities between
the properties of the \mypulsar\ system and those of several other known
binary pulsar systems, motivating the identification of a new class of
binary pulsars. We suggest that this new class could result from either: a
standard accretion scenario of a magnetar or a high-magnetic field pulsar;
common envelope evolution with a low-mass star and a neutron star, similar
to what is expected for ultra-compact X-ray binaries; or, accretion induced
collapse of a white dwarf. We also report the detection of a possible
$\textrm{K}^{\prime}=19.30(15)$ infrared counterpart at the position of the
pulsar, which is relatively bright if the companion is a helium white dwarf
at the nominal distance, and discuss its implications for the pulsar's
companion and evolutionary history.
\end{abstract}


\section{Introduction} \label{s:intro}

A mid-Galactic latitude pulsar survey with the Parkes Radio Telescope
\citep{cra06} detected three new pulsars in binary systems, none of which
easily fits within the standard evolutionary scenarios proposed for the
majority of recycled pulsars. One of them, \MyPulsar, was independently
discovered during the reprocessing of the Parkes Multibeam Pulsar Survey
data \citep{fau04}. This 172-ms binary pulsar has a relatively high surface
dipole magnetic field strength ($B \equiv 3.2\times 10^{19} (P\dot P)^{1/2}
\, {\rm G}$ $= 1.7 \times 10^{10} \, {\rm G}$) suggesting it is mildly
recycled. However, it appears to have a very light companion (minimum mass
$0.085 \,\Msun$) in a tight and nearly circular 4.6-hr orbit (see Table
\ref{t:orbparams}). This type of orbit and companion are typical of those of
fully recycled pulsars (which we define as pulsars with $P\la 8$\,ms and
$B\la 10^9$\,G) with He white dwarf (WD) companions. Thus, why \MyPulsar\
escaped being fully recycled is a puzzle.

In addition to this atypical combination of spin and orbital properties,
\mypulsar\ exhibits strong pulsed radio flux modulations, making the pulsar
undetectable at 1400\,MHz for lengths of time ranging from a few tens of
seconds to tens of minutes. It has been suggested by \citet{fau04} that this
behavior might be the nulling phenomenon seen in a handful of slow, isolated
pulsars. Nulling is a broad-band, if not total, interruption of the radio
emission for a temporary period of time. On the other hand, although nulling
could affect pulsars in binary systems as well, many binary pulsars vary due
to external effects such as eclipses. Such external effects might explain
\mypulsar's variability as well.

In this paper, we first report on multi-frequency observations of this
pulsar that suggest the radio variability is not intrinsic to the pulsar.
However, our analysis does not show strong evidence of a correlation between
radio flux and orbital phase, as one might expect from traditional
eclipse-like variability. We then report on our infrared search for a
counterpart of the companion using the Canada-France-Hawaii Telescope
(CFHT). This observation has identified a $\textrm{K}^{\prime}=19.30(15)$
star at the position determined by the radio timing observations. Finally,
we examine why other properties of this pulsar make it incompatible with
standard evolutionary scenarios, and identify a few other systems which have
similar characteristics. The addition of \mypulsar\ as an extreme case among
this group motivates us to identify a possible new class of binary pulsars.
We propose several possible evolutionary channels which might produce
members of this class and explain how the nature of the companion to
\mypulsar\ could be used to constrain the origin of these systems.

\section{Pulsed Radio Flux Variability} \label{s:analysis}

\begin{figure*}
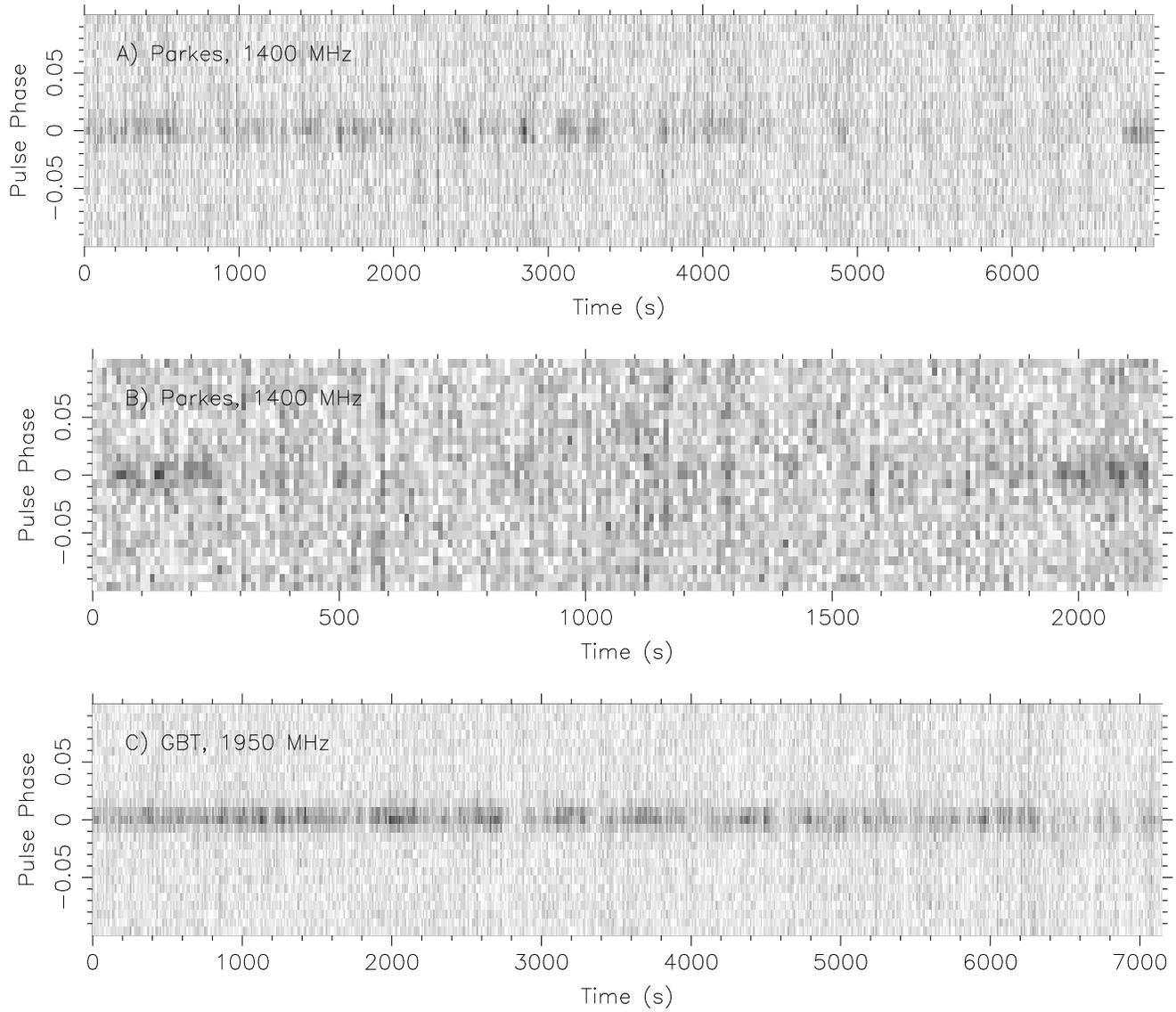

\centering \leavevmode
\includegraphics[angle=-90,scale=0.7]{f1a.ps}
\includegraphics[angle=-90,scale=0.7]{f1b.ps}
\includegraphics[angle=-90,scale=0.7]{f1c.ps}
\caption{Sample folded intensity profiles of \mypulsar\ as a function of
time for two 1400\,MHz observations at Parkes with 576\,MHz bandwidth (panel
A \& B) and a 1950\,MHz observation at GBT with 600\,MHz bandwidth (panel
C). The grayscale represents the intensity of the signal, with darker regions
being brighter. The center of the gaussian-like pulse profile should appear
at the pulse phase = 0.
\label{f:profile}}
\end{figure*}

The observed average pulsed radio emission from a pulsar can fluctuate for
several different reasons. These include effects from the pulsar itself, as
in nulling \citep[e.g.][]{bac70}, its environment, as in eclipsing binary
pulsars \citep[e.g.][]{fru88}, or the interstellar medium, as in
scintillation \citep[e.g.][]{ric70}. In the case of \mypulsar, during a
typical 1400\,MHz observation, the radio emission seems to turn on and off
randomly on timescales varying from tens of seconds to tens of minutes (see
Figure \ref{f:profile} for sample folded profiles). In many instances, the
pulsar is undetectable through entire observations ($\sim 15$\,min). In a
previous analysis of these radio flux modulations, \citet{fau04} concluded
on the basis of observations at 1400\,MHz that \mypulsar\ is probably a
pulsar experiencing pulse nulls. As nulling is a broad-band phenomenon
\citep[e.g.][]{bar81}, we decided to investigate the frequency dependence of
the fluctuations in order to test the nulling hypothesis.

\subsection{Data and Procedure} \label{ss:procedure}
The work we report is based on an extended dataset combining both the data
reported independently by \citet{fau04} and by \citet{ran06}. A total of 112
radio timing observations of \mypulsar\ were made at the 64-meter Parkes
Telescope and the 100-meter Green Bank Telescope between 2003 June and 2006
January. Relevant details for the current study are summarized in Table
\ref{t:setups} and we refer to the above two papers for more details about
the observational setups and timing results.

For the purpose of studying the radio emission variability, we made time
series of the pulsed flux intensity. We dedispersed the data at the pulsar's
dispersion measure (DM) of 148.5\,pc\,cm$^{-3}$ and then folded the
resulting time series in 10-s intervals using the timing ephemerides from
\citet{ran06}. For each observation, the pulse phase was determined from the
profile averaged over the entire observation. We fit each 10-s interval of
the folded pulse profile with a constant baseline plus a Gaussian of
variable amplitude having a fixed width at the predetermined phase. A
Gaussian FWHM=0.01964$P$, where $P$ is the pulse period, nicely fits the
profile averaged over many observations in the frequency range
680-4600\,MHz. Errors on the best-fit amplitudes returned by our
least-square minimization procedure were scaled under the assumption that
the off-pulse region RMS represents the total system noise. Although no flux
standard has been observed, we obtained pulsed flux density estimates by
using the radiometer equation and equating predicted noise levels to the
off-pulse RMS level. We also accounted for the offset between the telescope
pointing and the real position of the source (since in the early
observations the best-fit timing position had not yet been determined) by
approximating the telescope sensitivity to be an azimuthally symmetric
Gaussian having a FWHM corresponding to the radio telescope beam size, which
is 8.8$^{\prime}$ and 13.8$^{\prime}$ at 1400\,MHz for GBT and Parkes,
respectively.

\begin{figure*}
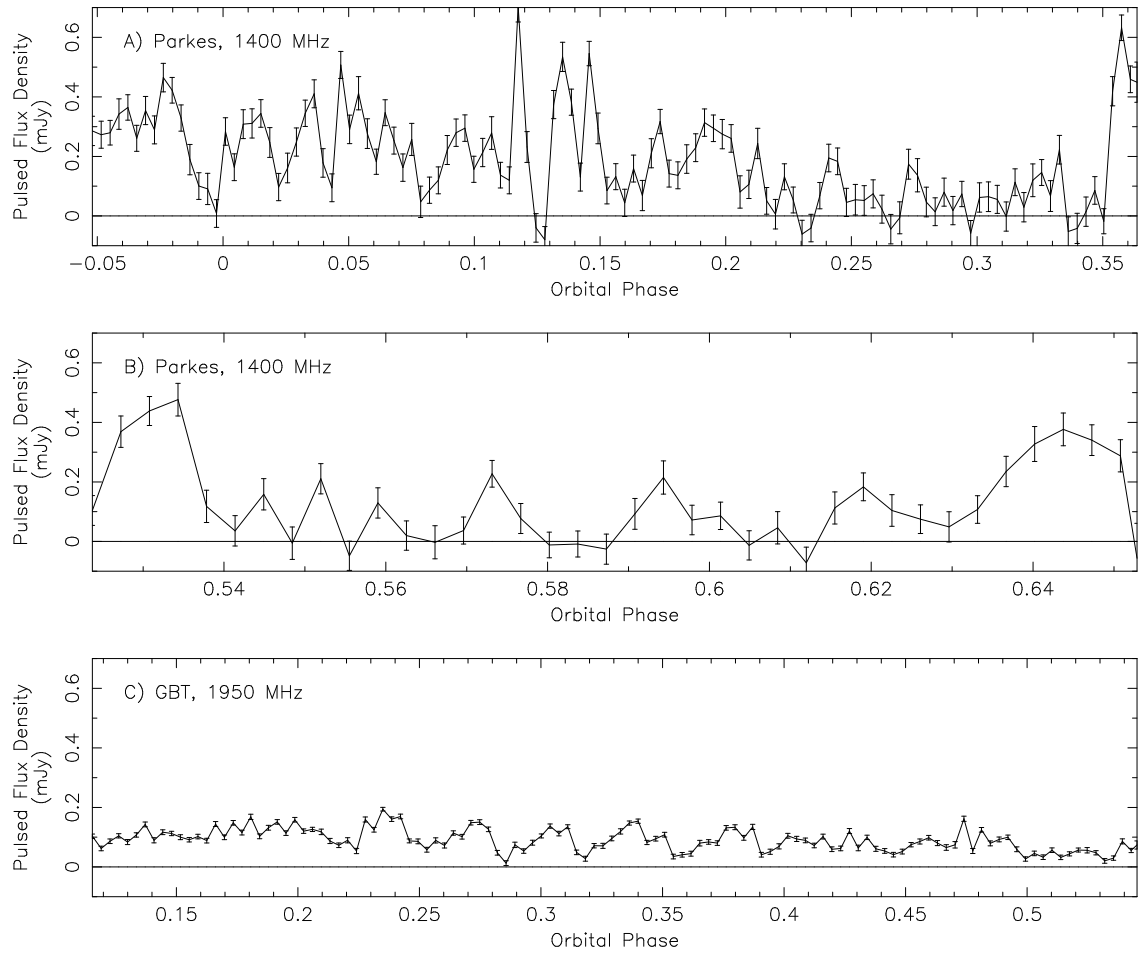

\centering
\leavevmode
\includegraphics[angle=-90,scale=0.6]{f2a.ps}
\includegraphics[angle=-90,scale=0.6]{f2b.ps}
\includegraphics[angle=-90,scale=0.6]{f2c.ps}
\caption{Sample light curves of the pulsed radio flux density as a
function of orbital phase. Each data point represents 60\,sec of data and
orbital phases are defined so that 0.25 is when the companion is in front of
the pulsar. Panel A, B and C are the two 1400\,MHz observations and the
1950\,MHz observation, respectively, displayed in Figure \ref{f:profile}. We
note that the flux drops below the detection limit at 1400\,MHz several
times whereas it appears to be always above this threshold at 1950\,MHz.
Also, as illustrated in Figure \ref{f:histogram}, the variability is much
stronger at lower frenquency.
\label{f:lightcurve}}
\end{figure*}

In this way, we generated flux time series for all 112 observations of
\mypulsar\ (see Figure \ref{f:lightcurve} for examples). As we describe
next, these results show that scintillation and nulling are highly unlikely
to be the origin of the observed variability.

\subsection{Radio-frequency-dependent Variability} \label{ss:nuller}
As \citet{fau04} discussed previously, interstellar scintillation (ISS) is
unlikely to be the source of fluctuations in \mypulsar. Scintillation is
produced by a diffractive scattering medium along our line of sight and the
typical diffractive scintillation timescale can be expressed as
\citep[see][]{cor98}:

\begin{equation}
   \Delta t_d = 2.53 \times 10^4 \frac{D \Delta \nu _d}{\nu V_{\rm ISS}}
                \,\, \textrm{s},
\end{equation}
with $D$ the distance to the source in kpc, $\Delta \nu _d$ the
decorrelation bandwidth in MHz, $\nu$ the observed frequency in GHz and
$V_{ISS}$ the velocity of ISS diffraction pattern in km\,s$^{-1}$. For the
1400\,MHz observation shown in Figure \ref{f:profile}, for example, the
NE2001 model for the Galactic distribution of free electrons \citep{cor02}
predicts $\Delta \nu_d = 0.01$\,MHz in the line of sight of \mypulsar\ at a
distance of 3.0\,kpc estimated from the DM. $V_{ISS}$ is typically dominated
by the pulsar velocities, which are in the range 10-100\,km\,s$^{-1}$ for
most binaries. Therefore, we estimate the scintillation timescale to be of
the order of a few seconds to a minute at most. This could be compatible
with the fast flux variations but can certainly not explain the extended
periods where the pulsar goes undetected. Perhaps most importantly, strong
scintillations will be averaged away since typical observing bandwidths are
much larger than the decorrelation bandwidth, and therefore contain many
``scintles''. Such averaging effectively rules out the ISS hypothesis.

Another possibility is that the flux modulation is related to intrinsic
nulling of the pulsar. Based on observations at 1400\,MHz only,
\citet{fau04} identified it as the most likely explanation. Only observed in
old, isolated pulsars so far, (though nothing prevents a binary pulsar from
being a nuller) nulling is a broad-band, if not total, interruption of the
radio emission \citep[e.g.][]{bar81}.

Considering the fraction of observations with no detection of radio emission
at various frequencies (see Table \ref{t:nodetection}), we note
qualitatively that \mypulsar\ is regularly undetectable at low frequencies
but easily detectable at high. For instance, we obtained four Parkes
observations at 680 and 2900\,MHz simultaneously in which the pulsar is
detected twice at 2900\,MHz while remaining undetected at 680\,MHz. In
addition, seven long GBT observations centered at 1850 and 1950\,MHz show
highly variable emission but little evidence that the pulsar ever disappears
completely (see Figure \ref{f:lightcurve}). Clearly, however, observations
could be biased by the relative instrumental sensitivity in each band and by
the intrinsic spectrum of the pulsar.

To investigate the effect of instrumental sensitivity and spectral energy
distribution, we analysed the pulsed flux densities at different
frequencies. Measured values were estimated using the radiometer equation as
explained in Section \ref{ss:procedure} and are displayed in Table
\ref{t:flux}. We note that \citet{fau04} reported a different flux density
than ours at 1400\,MHz (0.20(3) vs. 0.11(3)\,mJy, respectively). The
discrepancy could be because the average flux density changes depending on
the amount of time \mypulsar\ spends in its ``bright state'' during an
observation. The large standard deviation (0.16\,mJy, see Table
\ref{t:flux}) at this frequency suggests that by restricting the calculation
to observations for which \mypulsar\ is nicely detected, a higher flux value
can be obtained.

For the simultaneous observations at 680 and 2900\,MHz in which the pulsar
was not detected at 680\,MHz (see Table \ref{t:setups}), we can put an
interesting approximate lower limit on the spectral index, $\alpha$, of the
pulsar if we assume that the minimum detectable pulsed flux at 680\,MHz is
an upper limit to the pulsed flux at this frequency:

\begin{equation} \label{e:sensitivity}
   \alpha \gtrsim \frac{\log (S_{2900} / S_{680})}{\log (\nu _{2900} / \nu
   _{680})}
      \simeq 0.17 \,\, .
      \end{equation}

Such a value is extremely flat compared to that of the average population of
pulsars, which has a spectral index of $-1.8 \pm 0.2$ \citep{mar00}. Thus
either \mypulsar\ has a spectrum very different from those of most pulsars
and/or the flux variability is intrinsically frequency dependent.

\begin{figure*}
\centering \leavevmode
\includegraphics[angle=-90,scale=0.6]{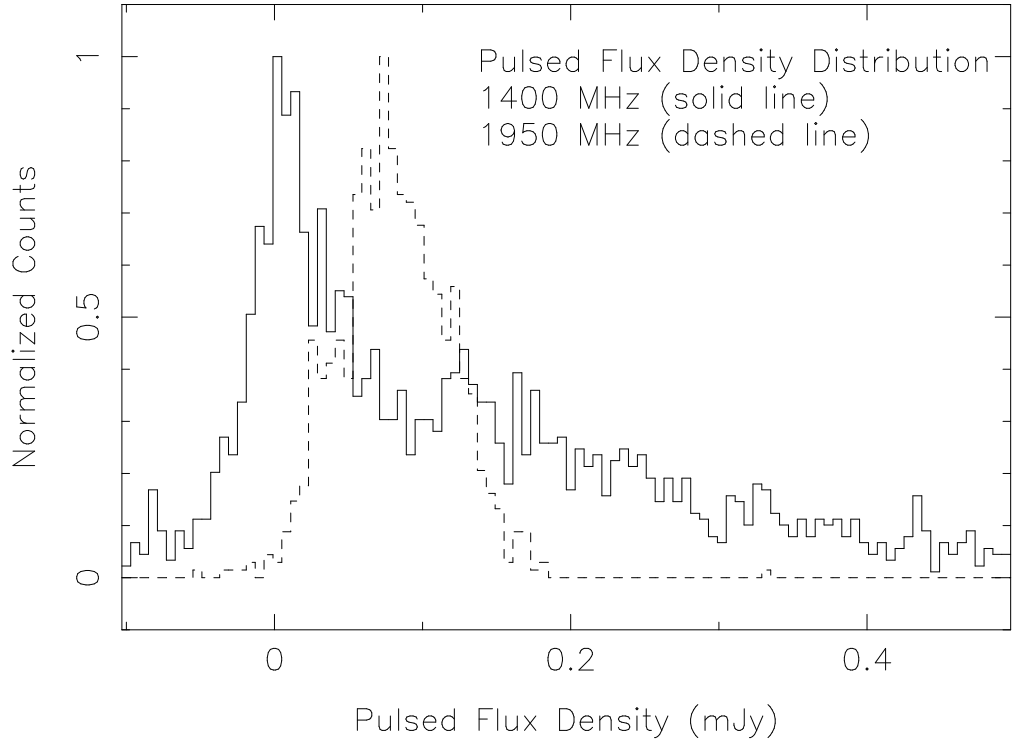}
\caption{Distribution of the measured pulsed flux density values at
1400\,MHz (solid line) and at 1950\,MHz (dashed line). The $3\sigma$
sensitivity threshold is $\sim$ 0.02 and 0.01\,mJy at 1400 and 1950\,MHz,
respectively. We observe that \mypulsar\ rarely disappears at 1950\,MHz
whereas there is a significant number of non-detections, centered about
zero, at 1400\,MHz (negative values are reported when the pulsed flux
density is below the telescope sensitivity, meaning the flux determination
algorithm has fit noise). Also, we note that the pulsed flux density is more
variable and spans over higher values at 1400\,MHz than at 1950\,MHz.
\label{f:histogram}}
\end{figure*}

The distribution of pulsed flux density values for all 1400 and 1950\,MHz
observations is shown in Figure \ref{f:histogram}. It is clear from the
distribution at 1400\,MHz that the numerous non-detections are responsible
for the peak below the sensitivity threshold. We also observe that the
1400\,MHz flux density has a higher average value and is much more variable
than at 1950\,MHz. This frequency dependence suggests some sort of
scattering mechanism with the unscattered flux level much higher than the
observed average flux at 1400\,MHz, and argues against it being classical
nulling. Therefore, we conclude that this unknown mechanism affecting the
lower frequency flux might be responsible for the apparent flat spectrum
derived from the simultaneous 680-2900\,MHz observation.

Recent simultaneous multi-frequency observations of PSR~B1133+16, a
well-established nuller, show that single pulse nulls do not always happen
simultaneously at all frequencies \citep{bha06}. They also observe, however,
that the overall null fraction does not present any evidence of frequency
dependence, which might mean that sometimes nulls are simply delayed at some
frequencies. In the case of \mypulsar, the S/N limits us to consider the
pulsed flux over times corresponding to many pulse periods only. Therefore,
the kind of non-simultaneous, frequency-dependent effect seen by
\citet{bha06} is not relevant to our analysis and thus we expect the
variability to be independent of frequency if really caused by nulling.

Although our flux measurements at other frequencies are not simultaneous, we
can assume they are good statistical estimates of the normal flux of the
pulsar and use them to characterize its spectrum. In an attempt to estimate
an unbiased spectral index, we can use the approximate maximum flux value at
each frequency. The 1400, 1850, 1950, 2900 and 4600\,MHz data give a
spectral index lying between $-1.5$ and $-3.0$, which is similar to many
known pulsars. However, it seems that the flux at 820\,MHz is much smaller
than expected from a single power-law spectrum. Since flux variations are
very important at low frequency and we only have a single detection at
820\,MHz, the reported maximum value is probably not representative of the
real flux of the pulsar at this frequency.

In summary, the facts that the pulsar radio emission rarely drops below our
detection threshold at 1950\,MHz and that the radio variability is frequency
dependent, demonstrate that the flux modulation is probably not classical
nulling. A non-nulling origin for the fluctuations at 1400\,MHz also
explains why \mypulsar\ does not fit in with expectations based on the
correlations observed between null fraction and spin period \citep{big92},
and between null fraction and equivalent pulse width \citep{li95}. In
comparison with nullers, it has one of the smallest spin periods and a small
pulse width ($\sim 3.4$\,ms), but one of the largest ``null'' fractions
($\sim 60$\% at 1400\,MHz). Since this pulsar is in a tight binary system,
the possibility of influence by its companion is therefore an important
alternative to consider.

\subsection{Orbital Correlation Analysis} \label{ss:orbital}
Even though a quick examination of the time series confirms that the flux
decreases observed for \mypulsar\ are not due to systematic eclipses of the
pulsar by its companion, a more subtle orbital correlation could exist. To
search for such an effect, we ask whether or not the pulsar is more likely
to be detected at a particular orbital phase. For this analysis, we folded
the time series in 1-min intervals, and defined the pulse as detected if the
best-fit Gaussian amplitude was greater than its 1$\sigma$ uncertainty. In
order to limit spectral effects, we restrict the analysis to observations
covering the range 1237.5--1516.5\,MHz at Parkes\footnote{This is the common
range of the observing modes centered at 1375 and 1400\,MHz.} and
1404.5--1497.5\,MHz at GBT.

\begin{figure*}
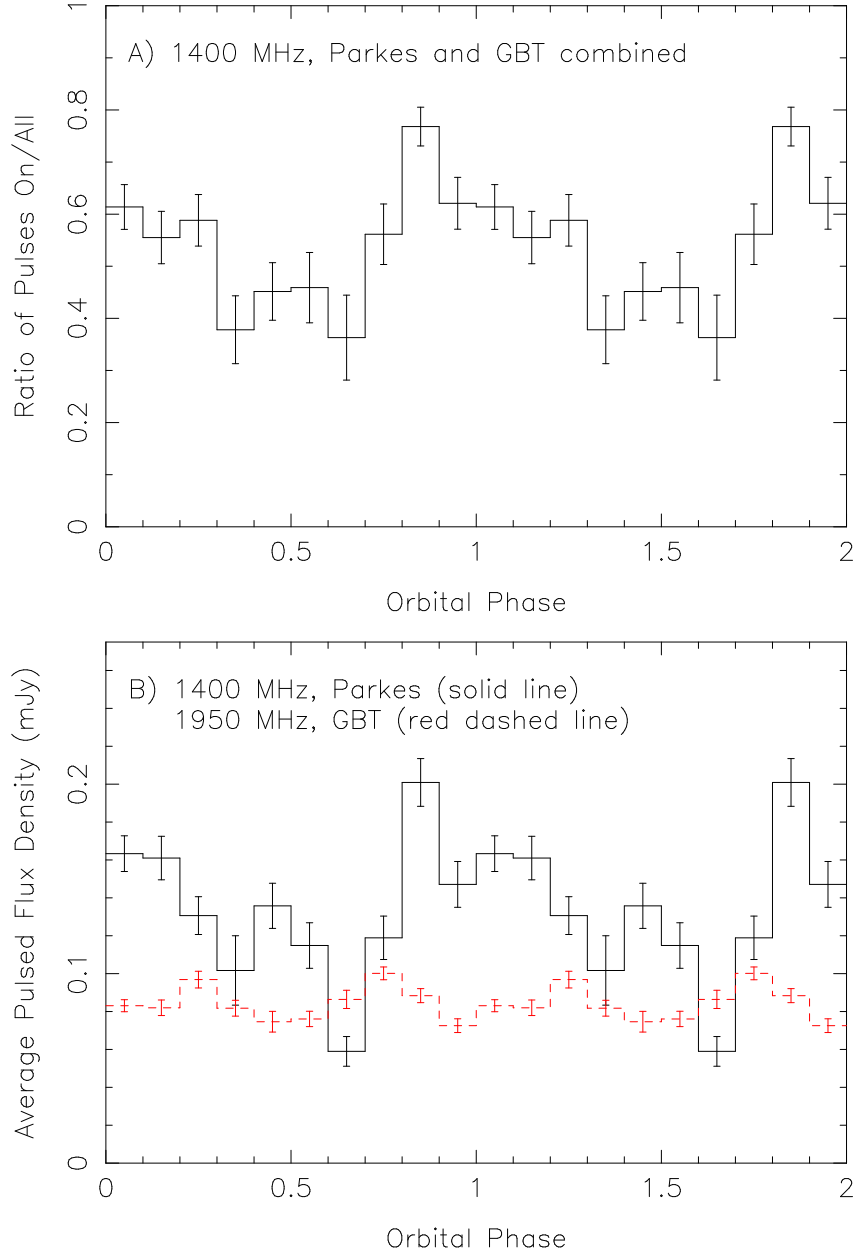

\centering
\leavevmode
\includegraphics[angle=-90,scale=0.5]{f4a.ps}
\includegraphics[angle=-90,scale=0.5]{f4b.ps}
\caption{Ratio of pulses detected with respect to the total number observed
at different orbital phases for the 1400\,MHz Parkes and GBT combined (panel
A) and a similar plot showing the average pulsed flux intensity (panel B)
for the Parkes 1400\,MHz data (solid line) and the GBT 1950\,MHz data (red
dashed line). Orbital phases are defined so that 0.25 is the pulsar's superior
conjunction (i.e. the companion passes in front of the pulsar). The scale of
the error bars represents the estimated errors without scaling from the
results of the Monte Carlo simulations, hence they are likely underestimated.
\label{f:analysis}}
\end{figure*}

The results of this analysis are shown in Figure \ref{f:analysis}a.  The
histogram represents the fraction of detected pulses with respect to the
total number observed, as a function of orbital phase. Errors were
determined using Poisson statistics, implicitly assuming our assigning of
each interval as a detection or non-detection is accurate.  There is a
suggestion that PSR J1744$-$3922 is more often undetectable between phases
$\sim 0.05-0.45$. The best-fit constant model gives a $\chi^2 /9 =5.85$
(the histogram has 10 orbital phase bins), which, if correct, would be
highly significant. To test the accuracy of our errors, we performed a Monte
Carlo simulation consisting of 10000 trials where we assigned each
measurement a random orbital phase. The mean $\chi^2 /9 =0.37$ with a
standard deviation of 0.18, suggesting our error estimates are significantly
overestimated.

To investigate further, we performed an analysis similar to the previous one
but for the pulsed flux density measured at each orbital phase averaged over
all observations. We selected two subsets of data: the Parkes and GBT
observations made at 1400 MHz and the GBT 1950 MHz observations. The first
one includes 101 observations which are on average $\sim 15$ minutes long,
but there are a few observations which are significantly longer. The latter
subset includes four observations, two of which have full orbital coverage,
one covering $\sim 75\%$ of the orbit and the last one $\sim 40\%$. 
Observations with no detection of \mypulsar\ are assigned upper limit values
of three times the background noise level (which is very small compared with
the average pulse of the pulsar when it is on). Errors in each bin of the
histogram are estimated from the RMS of the individual values in each
orbital bin. Results are plotted in Figure \ref{f:analysis}b. For the 1400
MHz data, a fit to a constant line has a $\chi^2 /9 =15.58$, and for the
GBT 1950\,MHz data, we find a $\chi^2 /9 =5.59$. Randomizing the
individual data points and folding them resulted in a $\chi^2 /9 = 1.02$
at both frequencies, suggesting our error estimates are reasonable. Although
this analysis strongly rules out the constant model for our folded light
curves, the shapes of these curves at 1400 and 1950\,MHz are not consistent
with each other.

This could be a result of the paucity of observations (typically 3--5) at
any given orbital phase. Therefore, random fluctuations in the flux on
timescales of tens of minutes (which we see in the time series) would likely
not be averaged out. To test whether or not the significant deviation from a
constant model depends on the particular phasing of our observations, we
again performed Monte Carlo simulations in which 10000 trial histograms
similar to those shown in Figure \ref{f:analysis} were generated from the
real data by adding a random orbital phase shift at the beginning of each
time series and then determining the $\chi^2 /9$ value for a constant
model. From the resulting distribution of $\chi^2 /9$ values, we estimate
the chance probability of obtaining the particular $\chi^2 /9$ values
obtained, or higher, using the real orbital phases. For the on-off analysis,
the probability is 0.102, or a formal $\chi^2 /9 =1.63$. For the 1400 and
1950\,MHz flux density light curves, the probabilities are 0.015 ($\chi^2
/9 =2.28$) and 0.212 ($\chi^2 /9 =1.34$). This suggests there may be some
correlation with the orbit, but the shape of our folded light curves are
still dominated by more stochastic flux variations given our limited data
set. We would expect a standard eclipse mechanism to make the pulsar dimmer
when the companion is in front of it at phase 0.25, which does not seem to
be the case. There may be large orbit to orbit variations in the eclipse
depths, durations, and phases. This kind of behavior has been observed in
other binary pulsar systems such as Ter5A, Ter5P, Ter5ad and NGC6397A
\citep[see][for examples]{ran05,hes06,dam01}, but a much larger data set
would be required to obtain a reliable average light curve to show if this
is the case for \mypulsar. It is still possible that the short time scale
pulsed flux variations tend to group in ``events" during which the pulse
gets dimmer. These ``events" may last for a significant fraction of the
orbital phase causing the apparent marginal orbital correlation given our
limited statistics.

\subsection{Accretion and mass loss limits} \label{ss:eclipse}
Many other systems are known to exhibit strong flux radio variations for
which the orbital dependence is well established. One of them is
PSR~B1718$-$19 \citep{lyn93}. Interestingly, this pulsar has a low-mass
companion and orbital properties similar to those of \mypulsar\ and is also
harder to detect at low frequency. At 408 and 606\,MHz, PSR~B1718$-$19 gets
so dim that it is barely detectable in spite of good sensitivity during a
large part of the orbit. On the other hand, at 1404 and 1660\,MHz the
orbital modulation of the average flux is much less important. This
flickering is probably made by material left over by the wind of the
companion, a bloated main sequence (MS) star \citep{jan05}. Although the
companion of PSR~B1718$-$19 is not large enough to come near to filling its
Roche-lobe, this could happen in a tighter binary system like that of
\mypulsar. In this case, some kind of tidal stripping could be occuring,
leaving material around the system. This could explain why the pulsar does
not disappear at conjunction like PSR~B1718$-$19 does, but in a more
stochastic way.

If the companion is losing mass, one might expect to observe small changes
in the orbital parameters. From radio timing \citep[see][]{ran06}, we can
set an upper limit of $|\dot P_{orb}| < 2 \times 10^{-10}$\,s\,s$^{-1}$ and
$|\dot x| < 7 \times 10^{-12}$\,lt-s\,s$^{-1}$ on the rate of change of the
orbital period and the rate of change of the projected semi-major axis,
respectively. We can use the latter quantity to evaluate the implied mass
loss limit. For a circular orbit, which is a good approximation here, we can
express the rate of change of the semi-major axis as \citep{ver93}:

\begin{equation}
   \frac{\dot a}{a} = 2 \frac{\dot J}{J} - 2 \frac{\dot M_{c}}{M_{c}}
                      \left(
		      1 - \frac{\beta M_{c}}{M_{p}} - 
		      \frac{(1-\beta) M_{c}}{2 (M_{p} + M_{c})} -
		      \alpha (1-\beta) \frac{M_{p}}{M_{p}+M_{c}} 
		      \right) \,\, ,
\end{equation}
where $x = a \sin i$, $\dot{x} = \dot{a} \sin i$, $M_{c}$ and $M_{p}$ are
the mass of the companion and the pulsar, respectively, $J$ is the total
angular momentum of the system, $\beta$ is the fraction of mass accreted by
the pulsar and $\alpha$ is the specific angular momentum of the mass lost in
units of the companion star's specific angular momentum.

For the case in which the total orbital angular momentum of the system is
preserved ($\dot J = 0$) we can see that mass loss from the
companion\footnote{Here we implicitly assume that $M_{c}<M_{p}$.} ($\dot
M_{c} < 0$) would necessarily lead to a widening ($\dot a/a > 0$) of the
orbit if: 1) the mass transfer is conservative ($\beta = 1$), or 2) the mass
transfer is non-conservative ($\beta < 1$) and $\alpha < 1+M_{c}/(2M_{p})$
\citep[see][for more details]{ver93}.

By considering the conservative case, in which $|\dot M_{c}|=|\dot M_{p}|$,
we obtain an upper limit on a possible mass accretion rate by the pulsar
$|\dot M_{p}| \lesssim 3 \times 10^{-12} \,\Msun\,{\rm yr}^{-1}$. This would
be even lower for the non-conservative case. Accretion onto the pulsar is
possible if the corotation radius:

\begin{equation}
   r_{co} \simeq 500 P_{{\rm psr}}^{2/3} M_{1.4}^{1/3} \quad \textrm{km},
\end{equation}
is larger than the magnetospheric radius of the pulsar:

\begin{equation}
   r_{mag} \simeq 800 \left( \frac{B_{{\rm psr}}^{4} R^{12}_{10}}{M_{1.4}
             \dot M^2_{-12}} \right)^{1/7} \quad \textrm{km},
\end{equation}
where $P_{{\rm psr}}$ and $B_{{\rm psr}}$ are, respectively, the spin period
and the surface dipole magnetic field in units of \mypulsar's $P$ and $B$
(172\,ms and $1.68 \times 10^{10}$\,G); $M_{1.4}$ is the mass of the pulsar
in units of $1.4\,\Msun$; $R_{10}$ is the radius of the pulsar in units of
10\,km; and $\dot M_{-12}$ is the accretion rate in units of
$10^{-12}$\,\Msun\,yr$^{-1}$. Given the upper limit on the mass accretion
rate, and the likely conservative $R_{10}=1$ value for the neutron star
radius, we find that $r_{mag}$ ($\gtrsim 600$\,km) is likely larger than
$r_{co}$ ($\sim 500$\,km). This argues against any significant accretion
occuring in the system.

That no significant accretion is occuring is also supported by an XMM-Newton
observation that allows \citet{ran06} to put a conservative upper limit
$\sim 2 \times 10^{31}$\,erg\,s$^{-1}$ on the unabsorbed X-ray flux from
accretion in the 0.1-10\,keV range. Assuming

\begin{equation}
   L_{X} = \frac{\eta G M \dot{M}}{R} \, ,
\end{equation}
with a conversion efficiency of accretion energy into observed X-rays
$\eta=0.1$, we get $\dot M_{p} \lesssim 2 \times 10^{-14} \, \Msun \,
\textrm{yr}^{-1}$ for accretion at the surface of the pulsar and $\dot M_{p}
\lesssim 2.4 \times 10^{-12} \, \Msun \, \textrm{yr}^{-1}$ if accretion is
limited to the boundary of the magnetospheric radius. Therefore, it appears
unlikely that \mypulsar\ is accreting and, if the companion is losing mass,
it is probably expelled away from the system.

\section{Infrared Observations} \label{s:infrared}

The radio flux variability, which might be due to material leaving the
surface of the companion, and the atypical evolutionary characteristics of
the pulsar (see Section \ref{s:evolution}) can be investigated further by
observing its companion at optical or near-infrared wavelengths. We imaged
the field of \mypulsar\ at $\textrm{K}^{\prime}$-band on the night of 2005
April 19 using the Canada-France-Hawaii 3.6-m Telescope (CFHT) at Mauna Kea.
The telescope is equipped with the Adaptive Optic Bonette
\citep[AOB,][]{rig98}, which provides good correction for atmospheric
seeing, and KIR, the 1024$\times$1024 pixel HAWAII infrared detector with
0\farcs0348 pixel scale. The total integration time was 30\,s$\times$59
integrations = 1770\,s.

\begin{figure*}
\centering
\includegraphics[angle=0,scale=0.5]{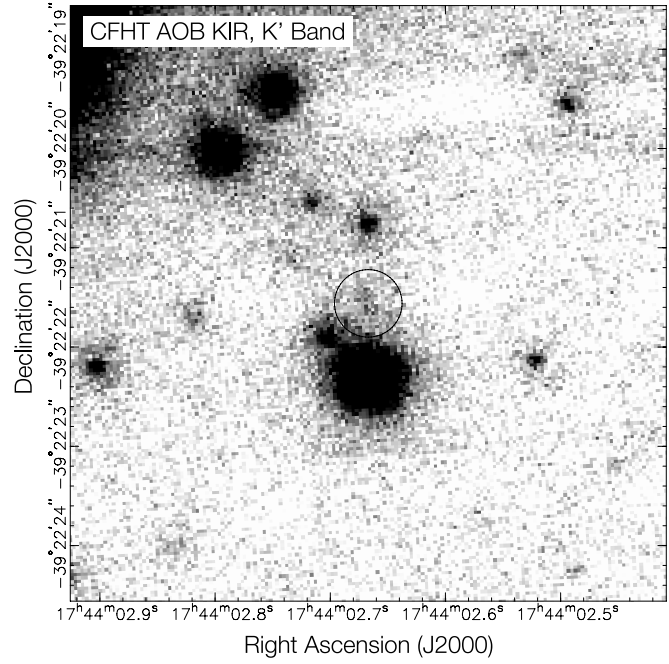}
\caption{Near-infrared image of the field of \mypulsar\ in the
$\textrm{K}^{\prime}$ band, obtained with AOB KIR at CFHT. The 0\farcs34
positional error circle ($3\sigma$ confidence) is shown, with the proposed
counterpart at the centre.
\label{f:IRimage}}
\end{figure*}

We substracted a dark frame from each image, and then constructed a
flat-field image from the median of the science frames. The images were then
flat-fielded, registered and stacked to make the final image. The final
stellar profile has a FWHM of 0\farcs17, degraded somewhat from the optimal
correction provided by the AOB system due to a high airmass ($\sim 2.0$) and
poor natural seeing. Figure \ref{f:IRimage} shows the final image we
obtained.

We analyzed the final CFHT image using the standard routine {\tt daophot}
\citep{ste87} for PSF fitting photometry, and calibrated the image using the
standard star FS34 \citep{cas92}. To find the photometric zero point, we
performed photometry on the standard star with a large aperture containing
most of the flux, and applied an aperture correction for the PSF stars in
the science image. Careful calibration of measurement errors has been done
by adding artificial stars of known magnitude to blank parts of the image
and then measuring their magnitude through the PSF fitting process along
with the real stars. Thus, errors on the magnitude returned by {\tt daophot}
can be rescaled by calculating the standard deviation for the added stars.

We found the astrometric solution for the image by cross-identifying five
stars with the 2MASS catalogue \citep{skr06}, fitting for scale, rotation
and displacement. The final astrometric uncertainty is 0\farcs34 at the
$3\sigma$ confidence level. This value depends on the matching to the
reference stars because the error on the radio timing position of \mypulsar\
is negligible, ($\sim 0\farcs03$). The final image is displayed in Figure
\ref{f:IRimage}, with the positional error circle centered at the radio
position of \mypulsar: $\alpha=17^{\rm h} 44^{\rm m} 02\fs 667(1)$ and
$\delta=-39\arcdeg 22\arcmin 21\farcs 52(5)$. Only one star falls inside
this circle, and for this we measure $\textrm{K}^{\prime}= 19.30(15)$. We
observe that, above the $3\sigma$ detection limit, the average stellar
density is 0.079\,arcsec$^{-2}$ and hence the probability of
a star falling in the error circle is only 2.9\,\%. Due to its positional
coincidence and the low probability of chance superposition, we henceforth
refer to this object as the possible counterpart to \mypulsar.

Unfortunately our near-infrared observation does not tightly constrain the
nature of the companion to \mypulsar, mainly because of the uncertainties in
the distance to the system and in the companion mass, as well as the fact
that its temperature is unknown. Assuming the NE2001 \citep{cor02} electron
density model is correct, we infer a DM distance of $3.0 \pm 0.6$\,kpc and
hence a distance modulus ranging from 11.9 to 12.8. Using the
three-dimensional Galactic extinction model of \citet{dri03} we get a value
of $A_{V} \simeq 1.9$ for a distance of 3.0\,kpc. Converting\footnote{For
simplicity, and because the error on the magnitude is dominated by the
distance estimate, we hereafter assume that K and $\textrm{K}^{\prime}$ are
similar.} the inferred extinction to $\textrm{K}^{\prime}$ band gives
$A_{K'} \simeq 0.21$ \citep[see][for conversion factors]{rie85}. Therefore
the estimated absolute magnitude of the counterpart lies in the range
$M_{K^{\prime}} \simeq [6.1,7.4]$.

We can evaluate how probable it is that the companion is a He WD, a typical
low-mass companion in binary pulsar systems, since WD cooling models can put
restrictions on the stellar mass and cooling age, given an observed flux.
Figure \ref{f:cooling} shows the absolute $\textrm{K}^{\prime}$ magnitude as
a function of cooling age for He WDs of different masses. These cooling
tracks were made by using WD atmosphere models based on the calculations of
\citet{ber95}, and thereafter improved by \citet{ber01b} and \citet{ber01a},
in combination with evolution sequences calculated by \citet{dri99}.
Although the mass range of models available to us does not go below
0.179\,\Msun, we can deduce that to be so luminous, a He WD would need to
have a very low mass and a cooling age much lower than the characteristic
age of the pulsar (1.7\,Gyr). Even if the mass were equal to the lower limit
of 0.08\,\Msun\ derived from radio timing, it seems unlikely that such a
companion could be as old as the pulsar characteristic age. Therefore, if
the companion is indeed a He WD, the pulsar's spin period must be close to
the equilibrium spin frequency it reached at the end of the recycling
process. In this case, its characteristic age is an overestimate of its true
age.

Another possibility is that the companion is not a He WD. If the companion
is a low-mass main sequence (MS) star, then the minimum mass required to
match the lower limit on the absolute $\textrm{K}^{\prime}$ flux is $\sim$
0.25\,\Msun, regardless of the pulsar age. Such a companion mass requires a
favorable face-on orbit but this cannot be ruled out from the near-infrared
and radio data. On the other hand, a lower-mass bloated MS star could be
equally as bright, so is also a possibility.

We cannot constrain the nature of the counterpart very well from a
measurement in a single near-infrared filter. Ideally, obtaining a spectrum
could yield: 1) a precise determination of the nature of the counterpart, 2)
orbital Doppler shift measurements of the spectral lines which can prove the
association as well as determine the mass ratio, 3) in the case of a white
dwarf, an estimate of its cooling age from spectral line fitting.

\section{Binary Pulsar Evolution} \label{s:evolution}

The sporadic radio emission from \mypulsar\ is not the only indication that
there is something unusual about the pulsar's interaction with its
companion. The pulsar is in a tight and low eccentricity ($e<0.001$) orbit
with an apparently very light companion having a minimum mass of
0.08\,\Msun. On the other hand it has a relatively large surface magnetic
field ($1.7\times 10^{10}$\,G) and an extremely long spin period (172\,ms)
compared with other binary pulsars having low-mass companions \citep[see,
e.g.,][for recent reviews]{sta04,vkk04}. These properties, along with the
relatively bright near-infrared counterpart of the companion, make it
unusual and suggest it evolved differently than most binary pulsars.

\begin{figure*}
\centering
\leavevmode
\includegraphics[angle=-90,scale=.5]{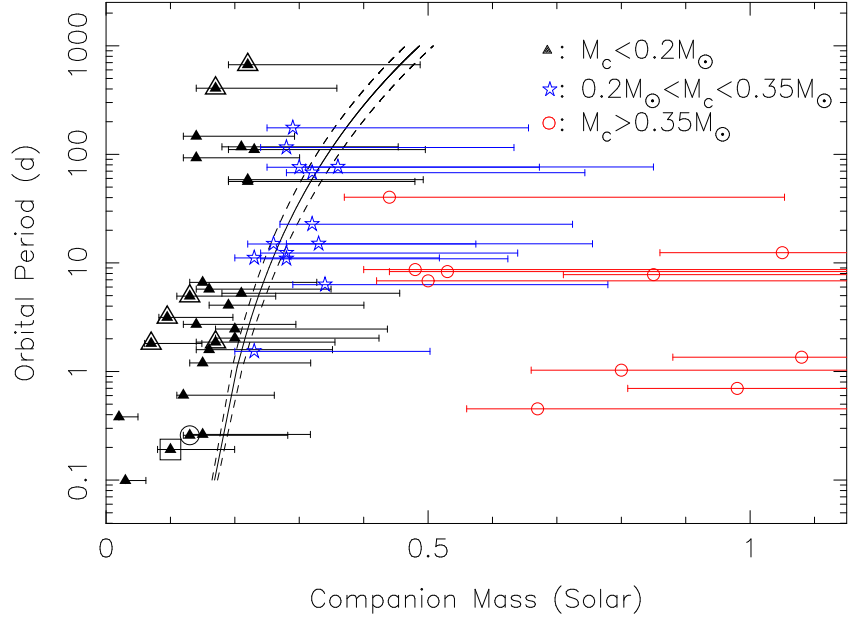}
\caption{Orbital period versus companion mass for binary pulsars in the
Galactic field in circular orbits ($e<0.01$). Symbols indicate the companion
median mass (corresponding to $i=60^{\circ}$) and are coded according to the
minimum mass: $M_{c}\leq 0.2 \,\Msun$, $0.2 \,\Msun < M_{c} \leq 0.35
\,\Msun$ and $M_{c} > 0.35 \,\Msun$ are black triangles, blue stars and red
circles, respectively. Error bars cover the 90\,\%-probability mass range
for randomly oriented orbits having $i=90^{\circ}$ to $i=26^{\circ}$. The
curves are the predicted $P_{orb}-M_{c}$ relationships for different
metallicity progenitors \citep[from][]{tau99}. \mypulsar\ is identified by a
square outline and pulsars listed in Table \ref{t:comparison} with triangle
outlines. The plot also includes the globular cluster pulsar PSR~B1718$-$19
(marked by a circle outline) because it resembles \mypulsar\ (see Section
\ref{ss:eclipse}). Data from the ATNF Pulsar Catalogue 
\citep{man05} (http://www.atnf.csiro.au/research/pulsar/psrcat/).
\label{f:mass_comp}}
\end{figure*}

In binary pulsar evolution, the nature of the companion plays a key role in
determining the final spin and orbital characteristics of the system. Most
of the binary pulsar population consists of pulsars with low-mass companions
in low-eccentricity orbits. Following the nomenclature adopted by
\citet{sta04}, the \emph{case A} evolutionary channel results in pulsars
having He WD companions. They are neutron stars (NS) which were spun up to
very short periods ($P \lesssim 8$\,ms) after accreting matter from a
low-mass star during a long and steady Roche-lobe overflow (RLO) phase
\citep{tau99}. For reasons that are not yet understood, this process is
responsible for the partial suppression of the surface magnetic field to
values of the order of $10^{8-9}$\,G. The most robust predictions of this
model are the correlations linking the orbital period to the mass of the He
WD \citep{rap95} (see Figure \ref{f:mass_comp}) and the eccentricity to the
orbital period \citep{phi92}.

On the other hand, the \emph{case B} channel is made of pulsars having more
massive CO WD or ONeMg WD companions ($M_{c} \gtrsim 0.45 \Msun$). Their
intermediate mass progenitors did not sustain a stable RLO phase, instead
evolving in a short-duration, non-conservative, common envelope (CE) phase
during which the pulsar spiraled into its companion's envelope. This process
only partly recycled the pulsar, leading to intermediate spin periods ($P
\gtrsim 8$\,ms) and leaving a higher magnetic field ($\sim 10^{9-10}$\,G).

\begin{figure*}
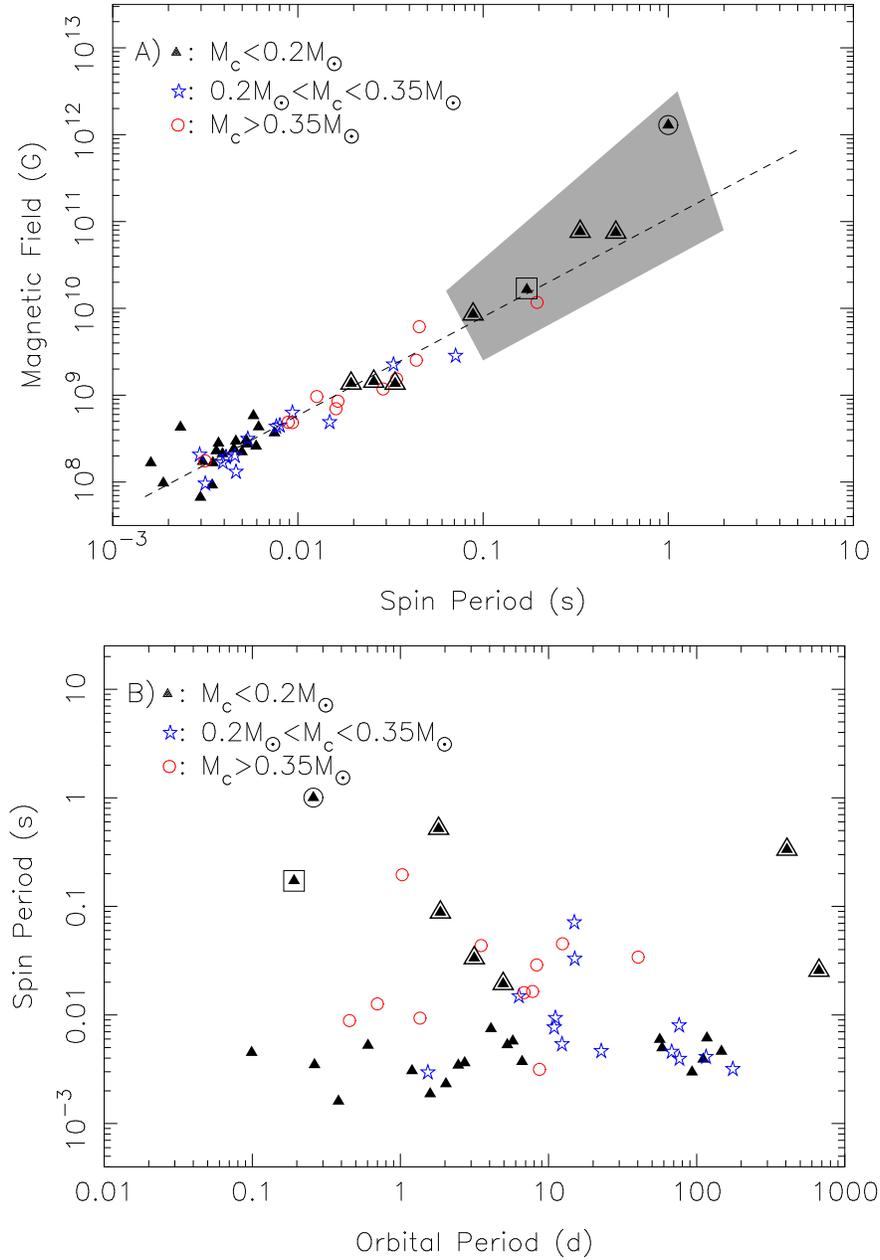

\centering \leavevmode
\includegraphics[angle=-90,scale=.5]{f8a.ps}
\includegraphics[angle=-90,scale=.5]{f8b.ps}
\caption{Inferred surface dipolar magnetic field strength versus spin period
(panel A) and spin period versus orbital period (panel B) for binary pulsars
in the Galactic field in circular orbits ($e<0.01$). Symbols are coded
according to the companion minimum mass: $M_{c}\leq 0.2 \,\Msun$, $0.2
\,\Msun < M_{c} \leq 0.35 \,\Msun$ and $M_{c} > 0.35 \,\Msun$ are black
triangles, blue stars and red circles, respectively. \mypulsar\ is
identified by a square outline and pulsars listed in Table
\ref{t:comparison} with triangle outlines. The shaded area is the
approximate region where is the proposed class of binary pulsars similar to
\mypulsar. The plots also include the globular cluster pulsar PSR~B1718$-$19
(marked by a circle outline) because it resembles \mypulsar\ (see Section
\ref{ss:eclipse}). The dashed line is the best-fit for a power-law, $B
\propto P^{\alpha}$, with $\alpha = 1.13$. We excluded PSR~B1718$-$19 from
the fit as it is in a globular cluster. Data from the ATNF Pulsar Catalogue
\citep{man05} (http://www.atnf.csiro.au/research/pulsar/psrcat/).
\label{f:recycling}}
\end{figure*}

In Table \ref{t:comparison}, the expected properties of systems resulting
from these two evolutionary channels are compared with those of \mypulsar.
Both scenarios fail to account for all the observed characteristics; this
suggests a special evolution for \mypulsar. This can also be seen from a
$P-B$ diagram (Figure \ref{f:recycling}a) made for binary pulsars in the
Galactic field having circular orbits. As opposed to isolated and other
kinds of non-recycled binary pulsars, there exists a very strong
relationship linking $P$ and $B$ which is presumably related to the
recycling process. To our knowledge, such a correlation has not been
reported in the recent literature although was indirectly found by
\citet{vdh95} who reported a possible correlation between $P-P_{orb}$ and
$B-P_{orb}$ for binary pulsars in circular orbits. The many binary pulsars
discovered in recent years may be making it easier to appreciate. In Figure
\ref{f:recycling}a we see that pulsars having light companions (\emph{case
A}) generally gather in the region of low magnetic field and short spin
period whereas the \emph{case B} pulsars lie in higher-valued regions.
Surprisingly, of the six highest magnetic field pulsars, five of them,
including \mypulsar, have light companions. The remaining one, PSR~B0655+64,
is an extreme system associated with the \emph{case B} channel since it has
a massive WD companion \citep{vkk04}. However, the \emph{case B} channel
cannot accommodate the other five pulsars because, assuming random orbital
inclinations, a simple statistical estimate gives less than a 0.1\,\%
probability for all of them to be more massive than the required $0.45 \,
\Msun$. For \mypulsar\ in particular, the orbital inclination would need to
be less than 12.5$^{\circ}$, which represents a 2.5\% probability.

We \emph{also} report in Table \ref{t:comparison} the principal
characteristics of binary pulsars that appear to be partly recycled (e.g.
$P>8$\,ms and $e<0.01$) but have companions likely not massive enough to be
explained by the standard \emph{case B} scenario. These pulsars have related
properties and could have experienced similar evolutionary histories. Apart
from their strange position in the $P-B$ diagram, they also stand out when
we look at the $P_{orb}-P$ relationship (Figure \ref{f:recycling}b). In this
plot, we see that pulsars having low-mass companions (\emph{case A}) occupy
the bottom region, below $P \lesssim 8$\,ms, and their spin periods are more
or less independent of the orbital period. This arises from the fact that
recycling probably saturates for a given accretion rate and/or accretion
mass \citep{kon04}. On the other hand, fewer constraints exist in this
parameter space for pulsars with massive WD companions (\emph{case B}).
Their short CE evolution would limit the recycling efficiency and thus the
final parameters are more sensitive to the initial conditions. Finally, we
observe a third category made of relatively slow pulsars with low-mass
companions in compact orbits. Neither the \emph{case A} nor the \emph{case
B} scenarios is able to explain such properties, especially for the most
extreme systems like \mypulsar\ and PSR~B1831$-$00. Therefore, we suggest a
new ``class'' of binary pulsars having the following properties: 1) long
spin periods (in comparison to MSPs), 2) large surface magnetic fields
($\sim 10^{10-11}$\,G), 3) low-mass companions, likely $0.08-0.3 \, \Msun$,
having nature yet to be determined, 4) low eccentricities, and possibly 5)
short orbital periods ($\lesssim 5$\,d). On this last point, very wide orbit
systems like PSR~B1800$-$27 and PSR~J0407+1607 might be explained by the
standard \emph{case A} scenario in which it is difficult to achieve an
extended period of mass transfer from the companion to the pulsar.

\section{Discussion} \label{s:discussion}

The existence of another ``class'' of binary pulsars is supported by the
fact that several pulsars now occupy a region of the parameter space
delimited by the spin period, orbital period, magnetic field and companion
mass that seems inaccessible to the standard evolutionary channels.
PSRs~J1744$-$3922, J1232$-$6501, B1718$-$19 and B1831$-$00 are certainly the
most noticeable candidates. Although other papers also identified that some
of these pulsars have strange characteristics \citep[see,
e.g.,][]{vdh95,sut00,edw01}, there is no consensus on their evolutionary
histories. For instance the case of PSR~B1718$-$19 was considered somewhat
unique because it is in a globular cluster and hence has possibly been
partly perturbed, or even greatly changed, by stellar interactions
\citep{erg96a}. The discovery of \mypulsar, in the Galactic field, is
important because it strengthens the possible connection between
PSR~B1718$-$19 and other similar binary pulsars in the field. Unless they do
not have WD companions, these pulsars were at least partially recycled
because tidal circularization is needed to explain eccentricities of 0.01
and smaller. However some of them have eccentricities $10^{-3}-10^{-2}$,
relatively large given their very short orbital periods, which is in
contrast to tight \emph{case A} systems having smaller eccentricity
\citep{phi92}. In this section we speculate and put constraints on some
scenarios that may explain this possible new class of pulsars.

\subsection{Recycled High Magnetic Field Pulsar Channel} \label{s:magnetar}
One possibility is that pulsars like \mypulsar\ may initially have had a
magnetar-strength magnetic field ($B\sim 10^{14}-10^{15}$\,G). Such a pulsar
could experience standard \emph{case A} evolution involving conservative RLO
but since the initial magnetic field is higher by 1--2 orders of magnitude,
at the end of the recycling process, the field might be $10^{10-11}$\,G
instead, consistent with observations. Since the recycling mechanism appears
to strongly correlate the final magnetic field with the final spin period,
the pulsar would also have an unusually long spin period as well.

Whether this extrapolation to magnetars and high magnetic field pulsars is
valid depends on accretion-induced decay models of the magnetic field
strength. Among the proposed mechanisms is magnetic field burial, in which
material accretes through the polar cap and, while piling up at the poles,
exerts a latitudinal pressure gradient by trying to spread toward the
equator. This effect tends to drag the field lines away from the poles,
increasing the polar cap radius and decreasing the magnetic moment of the
pulsar. \citet{pay04} show that the magnetic field would naturally
``freeze'' to a minimum stable strength once the amount of accreted mass
exceeds some critical value, if the magnetospheric radius is comparable to
the size of the neutron star. According to \citet{pay05}, this critical mass
could reach up to 1\,\Msun\ for a $10^{15}$\,G NS and hence magnetic field
suppression would be difficult to achieve due to the large amount of
accretion mass required. However, accounting for other effects like the
natural decay of magnetic field due to X-ray emission and high-energy bursts
\citep[see][for a review]{woo06}, might make a partially suppressed final
magnetic field of $~10^{10}$\,G plausible.

A recycled high magnetic field pulsar is expected to leave a He WD companion
and follow the companion mass-orbital period relationship as for the normal
\emph{case A} systems. PSR~B1718$-$19 is excluded from this kind of
evolution because its companion is a bloated MS star. All the other pulsars
listed in Table \ref{t:comparison} are potential members, albeit PSRs
B1800$-$27 and J0407+1607 would need relatively face-on orbits ($i < 30$ and
$24^\circ$, implying 13 and 9\% probability for randomly oriented orbits,
respectively) to match the $P_{orb}-M_{c}$ relationship (Fig.
\ref{f:mass_comp}). Another interesting prediction of this scenario is that,
because the recycling process can leave pulsars with relatively long spin
periods, they might not be very different from the spin periods when
accretion ceased. Thus, the real age could be much lower than the
timing-based characteristic age, and the WD companions would have younger
cooling ages as well.

The fraction of Galactic field binary pulsars in this class ($\sim$7/60)
might naively be thought to be similar to the fraction of observed magnetars
with respect to the ordinary pulsars ($\sim$10/1500). However, the observed
population of magnetars suffers from severe selection effects because many
of them appear to lie dormant, becoming observable for only brief intervals,
like XTE~J1810$-$197 \citep{ibr04}. Therefore, we can relate these
parameters as follows:

\begin{equation}
   \frac{N_{\mbox{\footnotesize class}}}{N_{\mbox{\footnotesize
   binary}}} = 
   \frac{N_{\mbox{\footnotesize obs. mag.}}}{N_{\mbox{\footnotesize radio}}}
   \frac{1}{f_{\mbox{\footnotesize quies}}} \,\, ,
\end{equation}
where $N_{\mbox{\footnotesize class}}$ is the number of pulsars in the new
class, $N_{\mbox{\footnotesize binary}}$ is the number of binary pulsar
systems, $N_{\mbox{\footnotesize obs. mag.}}$ is the number of observed
magnetars, $N_{\mbox{\footnotesize radio}}$ is the number of radio pulsars
and $f_{\mbox{\footnotesize quies}}$ the fraction of magnetars in
quiescence.

Hence we estimate that, due to quiescence, the fraction of observed
magnetars with respect to the total population is about
$\frac{10}{1500}/\frac{7}{60} \sim 0.06$. This would make a total population
of $\sim 175$ magnetars which is consistent with the possible $\sim 100$ in
our Galaxy estimated by \citet{woo06}, who used $f_{\mbox{\footnotesize
quies}} = 0.1$. Our crude calculation has many caveats: Is the binary
magnetar population similar to the binary radio pulsar population? Can the
short lifetime of magnetars and high magnetic field neutron stars limit the
number of such recycled systems? Clearly, this latter point depends strongly
on the time evolution of the self-induced decay of the magnetic field, which
seems to operate on a time scale of a few tens of thousands of years
\citep{kas04a}.

\subsection{UCXB Evolutionary Track Channel} \label{s:ucxb}
There exists a class of neutron stars and CO/ONeMg WDs that accrete from
light He WD donors in ultra tight (few tens of minutes) orbits. These X-ray
emitters, known as ultracompact X-ray binaries (UCXBs), may be the result of
wider $\sim 0.5$-day systems that have decayed due to gravitational wave
radiation. A CE scenario has been proposed as a viable channel for forming
CO/ONeMg WD -- He WD and NS -- He WD systems \citep{bel04}. In fact such a
channel would be very similar to the \emph{case B} scenario but for an
initially much lighter companion. For companions not massive enough to
experience the He flash, a CE phase is possible if the onset of mass
transfer occurs late enough in the evolution so that the companion has
reached the asymptotic giant branch. In this case, the RLO becomes unstable
and it bifurcates from the standard \emph{case A} track to the CE phase.

Pulsars experiencing this evolution would be partially recycled, like
\emph{case B} systems, but they would have He WD companions. After this
stage, only sufficiently tight systems with orbital periods of about one
hour or less can evolve to become UCXBs within a Hubble time because
gravitational decay is negligible for wider orbits. If \mypulsar-like
pulsars belong to the long orbital separation high-end of the UCXB formation
channel, we might expect to see more such pulsars at similar and smaller
orbital separations than \mypulsar. It is possible, however, that the
observed sample is biased: as the orbital separation decreases, wind and
mass loss by the companion become more important and would make them more
difficult to detect in classical pulsar surveys conducted at low frequency
(i.e. $\sim$400\,MHz) where eclipses are more frequent and radio emission
might simply turn off. The ongoing ALFA survey at Arecibo, at 1400\,MHz
\citep{cor06}, could therefore find several new pulsars like \mypulsar.
Additionally, larger orbital accelerations make them more difficult to find
and, since gravitational wave radiation varies as the fourth power of the
orbital separation, their lifetimes are dramatically shorter.

In such a scenario, these pulsars might be born at a spin period that is
comparable with those of \emph{case B} pulsars having CO WD companions,
assuming that the short, high-accretion rate recycling would efficiently
screen the magnetic field during the mass transfer process. Afterwards,
because of their relatively larger magnetic field, they would spin down more
rapidly than \emph{case B} pulsars. Thus, they would necessarily have longer
spin periods and the true age is more likely to be in agreement with the
measured characteristic age.

\subsection{AIC Channel}
Finally, a third scenario to explain the unusual properties of \mypulsar\ is
the accretion-induced collapse (AIC) of a massive ONeMg WD into a NS. Most
likely AIC progenitors would be massive ONeMg WDs ($\gtrsim 1.15 \, \Msun$)
accreting from MS donors in CV-like systems, or from red giants or He WDs in
UCXB systems \citep[see][for more details]{taa04,iva04}. \citet{nom91} have
shown that for accretion rates $\gtrsim 0.001 \, \dot{M}_{Edd}$ and/or
metal-rich accreted mass, the ONeMg WD would collapse to a NS rather than
explode in a supernova. More recent calculations including Coulomb
corrections to the equation of state by \citet{bra99} demonstrate that AIC
is possible for critical densities of the accreting WD core that are $30\%$
lower than previously found by \citet{nom91}, thus facilitating the
formation of neutron stars through this channel.

The properties predicted by the AIC scenario nicely agree with what we
observe for the class we are proposing: the mass transfer prior to the NS
formation would explain the low mass of the companion and the collapse is
expected to be a quiet event during which almost no mass is lost in the
system and only $\sim 0.2 \, \Msun$ is converted in binding energy into the
NS. This would keep the final orbital period close to what it was prior to
the collapse (which is small in most scenarios leading to AIC) and allow the
eccentricity to be very small or, at least, circularize rapidly. The
survival rate of such systems is probably higher than for standard systems
which are more likely to be disrupted if a large amount of mass is lost
during the supernova process. Although the initial properties of pulsars
formed by AIC are not known, we may presume they resemble those of
``normal'' pulsars with magnetic fields in the $10^{11-12}$\,G range.
Another interesting point to consider is that the AIC channel does not
require a degenerate companion. As opposed to the other two proposed
scenarios (Sections \ref{s:magnetar} and \ref{s:ucxb}), the formation of a
pulsar through AIC can interrupt the mass transfer, thus postponing further
evolution. If the ``evolutionary quiescence'' is long enough, we would
expect to find some non-degenerate companions around young AIC pulsars
having circular orbits but with spin periods and magnetic fields that are
more typical of isolated pulsars. Finally, as the companion continues to
evolve to fill its now larger Roche lobe, a shortened accretion phase might
occur, thus transfering a very small amount of mass to the pulsar and
leaving it partly recycled despite its low-mass companion.

Some binary pulsars among the group we highlighted, like PSRs B1831$-$00 and
B1718$-$19, have been proposed as AIC candidates in the past
\citep{vdh95,erg93}. However, there is no firm evidence to support this. For
instance, PSR~B1718$-$19 is presumably a member of the globular cluster
NGC~6342. As \citet{jan05} show, although observations suggest AIC is
possible, an encounter and tidal capture scenario cannot be ruled out and is
very reasonable given the plausible globular cluster association. On the
other hand, PSR~B1718$-$19 shares similarities with other binary pulsars in
the Galactic field, especially with \mypulsar. PSR~B1718$-$19's younger age,
larger magnetic field and spin period, as well as the fact that it has a
non-degenerate companion, are all compatible with it being an AIC pulsar in
the intermediate ``quiescent'' phase.

In this context, the other pulsars of our putative class would have reached
the final evolutionary stage and, hence, display mildly recycled properties.
If so, \mypulsar\ likely has a very light He WD companion. This might
explain the lack of traditional eclipses as in PSR~B1718$-$19 \citep{lyn93},
large DM variations as in NGC~6397A \citep{dam01}, and orbital period
derivatives as in other compact binary pulsars \citep{nic00} since
tidal effects are important for non-degenerate companions. If the residual
recycling phase left \mypulsar\ with a long spin period, the cooling age
of its hypothetical WD companion could be smaller than the characteristic
age of the pulsar for the same reason described in the recycled
high magnetic field pulsar scenario (see Section \ref{s:magnetar}). Also,
pulsars forming through AIC can have lower masses than those made in
standard supernovae and since they only accrete a small amount of mass from
their companion afterwards, they might be less massive than conventionally
formed MSPs.

\section{Conclusions} \label{s:conclusion}

This paper highlights the unusual nature of the binary pulsar system
\mypulsar. The puzzling radio flux modulation that it exhibits does not show
typical nulling properties as displayed by some old isolated pulsars;
specifically, we have found strong evidence that its variation is highly
frequency dependent. Although our orbital modulation analysis does not show
a significant correlation between orbital phase and flux, the modulation
could still be caused by a process related to a wind from its companion,
which results in short time scale variations grouped in extreme modulation
``events''. Additional monitoring of both the pulsar and of its companion
may prove useful in this regard.

We pointed out that this pulsar has an unusual combination of
characteristics: long spin period, very low-mass companion, high magnetic
field and short orbital period, that are unexplained by standard binary
pulsar evolution scenarios. We propose that \mypulsar, along with a several
other binary pulsar systems, are part of a new class of low-mass binary
pulsars which failed to be fully recyled. Specifically, we suggest three
alternative scenarios for this class of binary pulsars. Distinguishing among
them may be possible by improving our knowledge of the nature of their
companions. We also reported the detection of a possible near-infrared
counterpart to \mypulsar's companion, however, determining its nature will
require detailed near-infrared/optical follow-up.

\acknowledgments

R.P.B. would like to thank Carl Bignell for his help with the GBT
observations. Funding for this work was provided by NSERC Discovery Grant
Rgpin 228738-03, Fonds de Recherche de la Nature et des Technologies du
Qu\'ebec, the Canadian Institute for Advanced Research and Canada Foundation
for Innovation. V.M.K. is a Canada Research Chair. P.B. is a Cottrell
Scholar of Research Corporation. R.P.B. acknowledges the support from the
Student and Postdoc Observing Travel Fund of the National Research Council
of Canada's Herzberg Institute of Astrophysics. The Green Bank Telescope is
part of the National Radio Astronomy Observatory, a facility of the National
Science Foundation operated under cooperative agreement by Associated
Universities, Inc. The Parkes Observatory is operated by the Australia
Telescope National Facility, ATNF, a division of the Commonwealth Scientific
and Industrial Research Organisation, CSIRO.

\clearpage

\begin{deluxetable}{lc}
\tablecaption{Orbital and timing parameters for \mypulsar\tablenotemark{a}
\label{t:orbparams}}
\tablewidth{0pt}
\tablehead{\colhead{Parameter} & \colhead{Value}}
\startdata
Orbital Period, $P_{orb}$ (days)            \dotfill & 0.19140635(1) \\
Projected Semi-Major Axis, $x$ (lt-s) \dotfill & 0.21228(5) \\
Orbital Period Derivative, $|\dot P_{orb}|$ (s\,s$^{-1}$) \dotfill
& $<2 \times 10^{-10}$ \\
Projected Semi-Major Axis Derivative, $|\dot x|$ (lt-s\,s$^{-1}$) \dotfill
& $<7 \times 10^{-12}$ \\
\cutinhead{Derived Parameters}
Eccentricity, $e$                           \dotfill & $<$0.001 \\
Mass Function, $f_1$ (\Msun)                  \dotfill & 0.0002804(2) \\
Minimum Companion Mass, $M_c$ (\Msun)         \dotfill & $\geq$\,0.085 \\
Surface Dipole Magnetic Field, $B$ (G) \dotfill & $1.7 \times 10^{10}$ \\
Spin Down Energy Loss Rate (erg\,s$^{-1}$) \dotfill & $1.2 \times 10^{31}$ \\
Characteristic Age, (yr) \dotfill & $1.7 \times 10^9$ \\
\enddata
\tablecomments{Numbers in parentheses represent twice the formal errors in
the least significant digits as determined by {\tt TEMPO} after scaling the
TOA errors such that the reduced-$\chi^2$ of the fit was unity. The pulsar
is assumed to have mass 1.4\,\Msun.}
\tablenotetext{a}{Values reported from \citet{ran06}.}
\end{deluxetable}

\begin{deluxetable}{cccccccc}
\tablecaption{Details of the radio observations
\label{t:setups}}
\tablecolumns{8}
\tablewidth{0pt}
\tablehead{
\colhead{$\nu_{center}$\tablenotemark{a}} &
\colhead{BW\tablenotemark{b}} &
\colhead{Num. chan.\tablenotemark{c}} &
\colhead{Sampling} &
\colhead{$T_{sys}$\tablenotemark{d}} &
\colhead{Gain} &
\colhead{T\tablenotemark{e}} &
\colhead{Num. obs.\tablenotemark{f}} \\
\colhead{MHz}            & \colhead{MHz}               &
\colhead{}               & \colhead{$\mu$s}            &
\colhead{K}              & \colhead{K/Jy}              &
\colhead{min}            & \colhead{}                  }
\startdata
\multicolumn{8}{c}{Parkes Telescope} \\
\noalign{\vskip .8ex}
\hline
\noalign{\vskip .8ex}
680/2900\tablenotemark{g} & 56/576 & 256/192 & 500/500 & 68/31 & 0.625/0.59 & 20/20 & 5 \\
1375 & 288 & 96   & 500   & 28 & 0.71 & 17 & 69 \\
1400 & 576 & 192  & 250   & 32   & 0.71 & 15 & 13 \\
\cutinhead{Green Bank Telescope}
820  & 48  & 96   & 72    & 37   & 2.0  & 18  & 4  \\
1400 & 96  & 96   & 72    & 24   & 2.0  & 25  & 19 \\
1850 & 96  & 96   & 72    & 22   & 1.9  & 60  & 3  \\
1950 & 600 & 768  & 81.92 & 22   & 1.9  & 210 & 4  \\
4600 & 800 & 1024 & 81.92 & 20   & 1.85 & 257 & 1  \\
\enddata
\tablecomments{Receiver temperatures and gains are estimated operating
values. Parkes values are provided by J. Reynolds (private comm.). The system
temperature corresponds to the sum of the receiver temperature and the sky
temperature, which is determined from the 408\,MHz all-sky survey and
converted to other frequencies by assuming a power-law spectrum having a
spectral index $-2.6$ \citep{has82,has95}.}
\tablenotetext{a}{Central frequency of the receiver.}
\tablenotetext{b}{Observing bandwidth.}
\tablenotetext{c}{Number of frequency channels.}
\tablenotetext{d}{System temperature.}
\tablenotetext{e}{Average total integration time per observation.}
\tablenotetext{f}{Number of observations.}
\tablenotetext{g}{Observations were made simultaneously at these two
frequencies.}
\end{deluxetable}

\begin{deluxetable}{ccccccc}
\tablecaption{Percent of observations with no detection of \mypulsar
\label{t:nodetection}}
\tablecolumns{7}
\tablewidth{0pt}
\tablehead{\colhead{Frequency} &
\multicolumn{2}{l}{Parkes} &
\multicolumn{2}{l}{GBT} &
\multicolumn{2}{l}{Both} \\
\colhead{MHz} &
\colhead{\%} & \colhead{} & \colhead{\%} & \colhead{} &
\colhead{\%} & \colhead{}}
\startdata
680\tablenotemark{a}   & 100  & (5)  & --   & (--) & 100  & (5)   \\
820   & --   & (--) & 75 & (4)  & 75 & (4)   \\
1400  & 33 & (72) & 32 & (19) & 33 & (91)  \\
1850  & --   & (--) & 0  & (3)  & 0  & (3)   \\
1950  & --   & (--) & 0  & (4)  & 0  & (4)   \\
2900\tablenotemark{a}  & 50 & (4)  & --   & (--) & 50 & (4)   \\
4600  & --   & (--) & 0  & (1)  & 0  & (1)   \\
Total & 38 & (81) & 29 & (31) & 35 & (112) \\
\enddata
\tablecomments{Numbers in parentheses represent the total number of
observations for each band.}
\tablenotetext{a}{Observations at 680\,MHz and 2900\,MHz were simultaneous.
Excessive RFI contamination prevents us from using one of the 2900\,MHz
observations.}
\end{deluxetable}

\begin{deluxetable}{ccccc}
\tablecaption{Estimated Pulsed Flux Density of \mypulsar
\label{t:flux}}
\tablecolumns{5}
\tablewidth{0pt}
\tablehead{\colhead{Frequency} &
\colhead{Num. data\tablenotemark{a}} &
\colhead{Average} &
\colhead{Standard Deviation} &
\colhead{Maximum} \\
\colhead{MHz} & \colhead{} & \colhead{mJy} & \colhead{mJy} & \colhead{mJy}}
\startdata
680   &  40   & $<$0.07\tablenotemark{b} &  --  & --   \\
820   &  20   & 0.10  &  0.06  & 0.22 \\
1400  &  2244 & 0.11  &  0.16  & 0.45 \\
1850  &  142  & 0.11  &  0.04  & 0.20 \\
1950  &  852  & 0.08  &  0.04  & 0.19 \\
2900  &  40   & 0.09  &  0.08  & 0.16 \\
4600  &  258  & 0.006 &  0.012 & 0.03 \\
\enddata
\tablecomments{Pulsed flux density values were derived using the radiometer
equation, implicitly assuming that the off-pulse RMS is a good estimate of
the system temperature and that the sky emits according to the 408\,MHz
all-sky survey \citep{has82,has95}. Relative errors are estimated to be
$\sim 30$\,\%.}
\tablenotetext{a}{Number of data points used at each frequency. The time
resolution is 1 minute per data point.}
\tablenotetext{b}{$3\sigma$ upper limit.}
\end{deluxetable}

\begin{deluxetable}{lrrrrr}
\tablecaption{Characteristics of partly recycled binary pulsars
($P_{s}>8$\,ms) in the Galactic field in low-eccentricity orbits ($e<0.01$)
and having low-mass companions ($M_{c,min}<0.2 \, \Msun$)
\label{t:comparison}}
\tablecolumns{6} \tablewidth{0pt}
\tablehead{\colhead{Name} & \colhead{$P$} & \colhead{$\log{B}$} &
\colhead{$P_{orb}$} & \colhead{$M_{c,min}$\tablenotemark{a}} &
\colhead{Type} \\ \colhead{} & \colhead{ms} & \colhead{G} & \colhead{days} &
\colhead{\Msun} & \colhead{}} \startdata \emph{Case A} & $\lesssim 8$ & 8-9
& -- & $\lesssim 0.45$ & He WD \\ \emph{Case B} & $\gtrsim 8$ & 9-10 & -- &
$\gtrsim 0.45$ & CO WD \\ \noalign{\vskip .8ex} \hline \noalign{\vskip .8ex}
PSR~J1744$-$3922 & 172.44 & 10.22 & 0.19 & 0.08 & ? \\
PSR~B1718$-$19\tablenotemark{b} & 1004.03 & 12.11 & 0.25 & 0.12 & Bloated MS\tablenotemark{c} \\
PSR~B1831$-$00 & 520.95 & 10.87 & 1.81 & 0.06 & ? \\
PSR~J1232$-$6501 & 88.28 & 9.93 & 1.86 & 0.14 & ? \\
PSR~J1614$-$2318 & 33.50 & 9.14 & 3.15 & 0.08 & ? \\
PSR~J1745$-$0952 & 19.37 & 9.51 & 4.94 & 0.11 & ? \\
PSR~B1800$-$27 & 334.41 & 10.88 & 406.78 & 0.14 & ? \\
PSR~J0407+1607 & 25.70 & 9.16 & 669.07 & 0.19 & ?
\enddata
\tablenotetext{a}{$M_{c,min}$ refers to the minimum mass of the companion
corresponding to an orbital inclination angle of $90^\circ$ and assuming a
mass for the pulsar of $1.35 \, \Msun$. } \tablenotetext{b}{In globular
cluster NGC~6342.} \tablenotetext{c}{\citet{jan05}}
\end{deluxetable}

\clearpage

\begin{figure*}
\centering
\leavevmode
\includegraphics[angle=-90,scale=0.5]{f6.ps}
\caption{Cooling tracks for He WDs made using the evolution sequences of
\citet{dri99} and our atmosphere models \citep[see][for more
details]{ber95,ber01b,ber01a}. Lines show the cooling for constant masses of
0.179, 0.195, 0.234, 0.259, 0.300 and 0.331\,\Msun, from bottom to top,
respectively. The shaded region is the restricted range of absolute
$\textrm{K}^{\prime}$ band magnitude inferred from the CFHT data, and the
dotted vertical line indicates the characteristic age of \mypulsar.
\label{f:cooling}}
\end{figure*}

\end{document}